\documentclass[12pt]{article}
\usepackage[a4paper,margin=1in]{geometry}
\usepackage{csquotes}
\usepackage[style=apa,natbib=true,backend=biber]{biblatex}
\usepackage{amssymb}
\usepackage{amsmath}
\usepackage{graphicx}
\usepackage{hyperref}
\usepackage{booktabs}
\usepackage{multirow}
\usepackage{makecell}
\usepackage{siunitx}
\usepackage{threeparttable}
\usepackage{pdflscape}
\usepackage{adjustbox}
\usepackage{tabularx}
\usepackage{longtable}
\usepackage{ragged2e}
\usepackage{microtype}
\usepackage[section]{placeins}
\usepackage[english]{babel}

\newcolumntype{Y}{>{\RaggedRight\arraybackslash}X}

\begin{document}

\title{National-Scale Multimodal Accessibility Fields from Open Geospatial Data}
\author{%
  Till Zacher\textsuperscript{1,2,*} \quad Louis Stille-H\"onig\textsuperscript{1,2} \quad
  Markus Lienkamp\textsuperscript{1,2}\\[8pt]
  \parbox{\dimexpr\textwidth-2\tabcolsep\relax}{\centering\small
  \textsuperscript{1}Institute of Automotive Technology, Technical University of Munich,
  Garching, Germany\\
  \textsuperscript{2}Munich Institute of Robotics and Machine Intelligence (MIRMI),\\
  Technical University of Munich, Germany\\
  \textsuperscript{*}Corresponding author:
  \href{mailto:till.zacher@tum.de}{till.zacher@tum.de}}}
\date{}
\maketitle
\thispagestyle{plain}

\begin{abstract}
Large-area accessibility analysis remains difficult to reproduce because
multimodal routing, heterogeneous opportunity data, and reusable spatial
outputs are rarely integrated in a single transparent workflow. We present MAPPER (Multimodal Accessibility and Place Profiling
Engine for Arbitrary Regions), an open-data pipeline for multimodal
accessibility analytics. The workflow builds walk, bike, car, and public
transport travel-time graphs from OpenStreetMap and GTFS data, harmonizes
mapped POI- and proxy layers representing the land-use side of the
accessibility relation, and computes continuous accessibility fields using
network kernel density estimation. A low-overhead native compute core
(contraction-hierarchy routing, C++/OpenMP accessibility accumulation,
DAG-style orchestration eliminating redundant computation) is the enabler that
makes many graphs, large extents, and a broad layer set jointly tractable on
commodity hardware. Outputs are sampled to a uniform H3 grid for downstream
urban and regional analysis.

A Germany-wide production run yields an H3-indexed accessibility
dataset with nearly 40 million cells and 972 mode-layer fields across nine
network and intermodal modes. Routing comparisons against an empirical mobility
panel show low mean bias for walking, cycling, and car, while public
transport results are interpreted more cautiously as schedule-based supply
estimates sensitive to waiting and transfer assumptions. A Bavaria case study
operationalizes the travel-time targets of the state's central-place planning
system on the resulting fields and reveals strong modal differences in
accessibility to basic services.

The contribution is a reproducible framework for large-area multimodal
accessibility analysis, with an aligned output structure designed for reuse in
urban and regional analysis, planning support, and scenario modeling.
\end{abstract}


\section{Introduction}
\label{sec:introduction}

Accessibility measures represent how transport networks and the spatial
distribution of opportunities jointly shape the potential for interaction
across urban and regional systems \citep{hansen1959, geurs2004}.
Yet large-area accessibility analysis remains difficult to implement in practice due to large combinatorial complexity. Many studies
rely on bespoke preprocessing, single-mode routing, coarse zonal summaries, or
locally tailored data assemblies, which makes results harder to compare,
transfer, and reuse in downstream urban analytics \citep{marwal2022,
mazzulla2024}.

These limitations matter for transport and land-use analysis in particular,
because accessibility is the standard quantity through which the interaction
of the two systems is measured: the transport side enters through multimodal
travel-time impedance, and the land-use side enters through the spatial
distribution of opportunities: services, workplaces, and other activity
locations. Comparative questions about this interaction, such as whether
land-use plans presume levels of access that the transport system actually
delivers, require accessibility measures that are consistent across modes,
regions, and opportunity types. National and supraregional analysis therefore
requires workflows that can integrate multiple travel modes, heterogeneous
opportunity representations, and a common output structure while remaining
computationally tractable. The computational burden also grows quickly with
the product of transport modes, opportunity classes, spatial granularity, and
study extent. Continuous network-based fields are attractive in this setting
because they reduce dependence on administrative units, preserve local spatial
gradients, and produce smoothly varying surfaces that can be reused in
planning analysis, scenario studies, and downstream statistical modeling. But
they also require explicit choices about impedance modeling, source
harmonization, and scalable computation.

Common approaches such as cumulative opportunity measures and catchment-based
service-accessibility methods are useful for specific planning questions, but
they frequently produce threshold-dependent or zone-based summaries that are
less convenient to merge with other spatial layers \citep{santanapalacios2022, kim2018, lin2024}.
For open-data accessibility analysis to become reusable geospatial
infrastructure, workflows must connect multimodal travel-time graphs,
harmonized opportunity layers, and analysis-ready outputs at scale
\citep{nelson2019, fink2024}.

We present MAPPER (Multimodal Accessibility and Place Profiling Engine for
Arbitrary Regions), an open-data pipeline for large-area multimodal
accessibility analysis. MAPPER constructs walk, bike, car, and public
transport travel-time graphs from OpenStreetMap and GTFS schedules;
harmonizes mapped amenities, workplace proxies, and census-aligned covariates, the land-use side of the accessibility relation, onto a common
reference structure; computes continuous accessibility fields using network
kernel density estimation (NKDE); and samples the results to a uniform H3
grid for downstream urban and regional analysis. We demonstrate the
workflow for Germany and use a case study in Bavaria to show how the resulting
layers can support comparative assessment across modes. We interpret the
routing evaluation as a check on impedance realism rather than as
complete validation of accessibility inference.

\subsection{Contributions}
\label{subsec:contributions}

The contributions of this paper are:
\begin{itemize}
  \item a reproducible open-data pipeline for country-scale multimodal
  accessibility analysis using OSM-, GTFS-, and census-linked inputs, built on
  a low-overhead native compute core (contraction-hierarchy routing, C++/OpenMP
  accessibility accumulation, DAG-style orchestration) that makes national-scale
  runs feasible on commodity hardware;
  \item a harmonized NKDE-based workflow that generates continuous
  accessibility fields for more than 100 opportunity categories and exports
  them as dense H3-indexed multimodal feature stacks;
  \item an H3-aligned output representation designed for reuse in urban and
  regional analytics, comparative diagnostics, and planning applications such
  as service-provision assessment;
  \item an empirical demonstration for Germany that combines country-scale
  computational feasibility and routing diagnostics with a Bavaria case study
  operationalizing the travel-time targets of the state's central-place
  planning system on the resulting fields.
\end{itemize}

The remainder of the paper is organized as follows. Section~\ref{sec:related_work}
positions MAPPER against existing accessibility methods and large-area
geospatial workflows. Section~\ref{sec:data} describes the data
sources and their scope. Section~\ref{sec:methods} details graph construction,
source harmonization, NKDE, and H3 sampling. Section~\ref{sec:validation}
clarifies what the routing diagnostics validate. Section~\ref{sec:results}
reports the main empirical findings, followed by a discussion and conclusion.

\section{Related Work}
\label{sec:related_work}

\subsection{Accessibility measures and operational trade-offs}
\label{subsec:rw_accessibility}

Accessibility has long been framed as the potential to reach opportunities
under spatial or temporal impedance \citep{hansen1959, ingram1971}. Subsequent
reviews distinguish among cumulative-opportunity, gravity-based, utility-based,
and person-based formulations, each with different trade-offs in theoretical
richness, data requirements, and interpretability \citep{pirie1979, geurs2004,
kwan1998, marwal2022, mazzulla2024}. For planning practice, these trade-offs
matter because the chosen measure affects not only the numeric results but also
the kinds of comparisons that can be made across places and modes.

Cumulative opportunity measures remain attractive because they are easy to
communicate and operationalize, while gravity-style measures retain continuous
distance decay and can better represent gradual reductions in reachability
\citep{santanapalacios2022}; recent work operationalizes gravity-style,
multi-activity access into communicable open-source indices by applying
diminishing-returns transformations to the otherwise unbounded potential
\citep{roper2023}. Two-step floating catchment area variants add
supply-demand logic and are widely used in sector-specific applications such as
healthcare and greenspace accessibility, but they typically remain sensitive to
catchment definitions, threshold choices, and zonal aggregation
\citep{kim2018, lin2024}. Beyond the choice of measure, place-based
accessibility results are also tool-dependent: \citet{higgins2022} show that
identical origin--destination inputs produce diverging travel times and
accessibility surfaces across standard software platforms, with spatially
clustered differences --- an ``algorithmic dependence'' that argues for
workflows whose impedance model, decay function, and aggregation rules are
fully specified and reproducible rather than embedded in tool defaults.
Because this paper aims to produce reusable,
multi-layer accessibility fields rather than a single threshold-based planning
indicator, we position MAPPER within the gravity-style family and make the
trade-off explicit: we prioritize continuous, place-based surfaces over
person-specific or capacity-equilibrium formulations.

\subsection{Network-constrained continuous accessibility fields}
\label{subsec:rw_network}

When accessibility is computed on transport networks, network-based impedance
is generally preferable to straight-line proximity because it respects actual
reachability constraints. This is particularly important in multimodal and
infrastructure-constrained settings, where Euclidean smoothing can allocate
influence into unreachable space \citep{xie2008, okabe2009}. Network kernel
density estimation provides one route to continuous field construction on
networks, and prior methodological work has shown both its conceptual
advantages over planar KDE in network contexts and the sensitivity of results
to bandwidth and kernel choices \citep{xie2008, okabe2009, mcswiggan2017,
martinez2013}.

This literature is directly relevant to MAPPER because the paper does not seek
to introduce a new accessibility theory; it seeks to operationalize a network-
based, reusable accessibility workflow at large scale. The methodological
question is therefore not whether NKDE is universally superior to every other
measure, but whether it is an appropriate choice for generating aligned,
continuous accessibility layers from shared opportunity inventories across
multiple modes. In that narrower sense, prior NKDE and distance-decay work
provides the main conceptual basis for the approach adopted here
\citep{xie2008, okabe2009, martinez2013}. A practical consequence of the
continuous formulation matters for planning use: threshold and scenario
choices can be applied \emph{after} the field is computed, rather than being
baked into the measure beforehand as in cumulative-opportunity indicators.
The Bavaria application exercises exactly this property, and
Appendix~E reports how the resulting classifications
respond to threshold variation.

\subsection{Scale, multimodality, and reusable outputs}
\label{subsec:rw_scale}

Much applied accessibility work remains focused on single cities, single
service classes, or a limited set of transport modes. Examples include walk-to-public-transport access studies, sector-specific service accessibility analyses, and
modal-access-gap studies that are analytically valuable but intentionally
narrow in thematic scope \citep{garcia-palomares2013, higgs2015, maharjan2022}.
At the same time, a growing set of large-area accessibility datasets has shown
the feasibility of reproducible, gridded, travel-time-based analysis, including
global indicators, European city-level accessibility datasets, and regional
time-sensitive matrix products \citep{nelson2019, christodoulou2020, fink2024}.

These large-area products are important precedents, but they usually prioritize
travel-time matrices or a narrower indicator set rather than a broad stack of
continuous, multimodal opportunity fields. Open-source planning instruments and
multimodal workplace-accessibility workflows also demonstrate the practical
value of accessible tooling, yet they are typically more geographically or
thematically specific than the infrastructure-oriented framing pursued here
\citep{pajares2022, pfertner2023}. They are also generally delivered as
interpreter-level packages or containerized services, a design that favors ease
of adoption over raw computational throughput; at the mode--layer--extent
product targeted in this paper, the resulting orchestration overhead becomes a
binding constraint, which motivates the native-code architecture described in
Section~\ref{subsec:methods_overview}. For MAPPER, the relevant gap is therefore
not simply ``large scale'' or ``multimodal'' in isolation, but the combination
of scale, multimodality, broad opportunity coverage, and an output structure
that can be reused in downstream urban analytics.
Hierarchical discrete global grid systems (DGGS) are relevant in this context
because they provide a standards-driven way to index, aggregate, and join large
geospatial datasets on a common multi-resolution cell structure rather than
relying only on administrative zones or ad hoc rasterizations
\citep{gibba, bondaruk2020, kmoch2022}. Recent evaluations covering H3 show that
such grids are operational at large scale and support practical conversion and
integration workflows, while also making clear that interoperability and full
standards compliance remain incomplete in current implementations
\citep{bondaruk2020, kmoch2022}.

\subsection{Open geospatial opportunity data and schema design}
\label{subsec:rw_opportunity}

The other side of the problem is opportunity representation. OpenStreetMap has
become an increasingly useful source for activity locations and attraction
proxies in travel-demand and accessibility workflows, with prior work showing
that OSM-derived POIs and building information can support large-scale location
and capacity estimation when combined with explicit plausibility checks and
aggregation rules \citep{klinkhardt2021, malkus2024, notelaers2024}. This
supports the use of open-data opportunity inventories, but it also highlights a
recurring challenge: raw OSM categories are noisy, imbalanced, and not designed
for stable analytical schemas out of the box.

Related urban analytics work further shows that POIs carry meaningful signals
about urban function, while also underscoring the need to manage semantic
heterogeneity and class imbalance \citep{yang2022, yu2022}. In this context,
the main methodological burden is to justify a transparent and reproducible
opportunity schema built from mapped amenities and proxy layers. Recent work
also makes this burden more concrete: OSM POI quality and completeness vary by
category and can be assessed using large-scale external validation and
indicator-based diagnostics, while complementary extraction pipelines show how
heterogeneous OSM tags can be cleaned and categorized into stable POI classes
for reproducible downstream analysis \citep{klinkhardt2023, notelaers2024}.
Germany-wide POI generation workflows further demonstrate that such
harmonization can be carried through at national scale, while also revealing
remaining challenges such as multi-activity assignment and regionally varying
data quality \citep{malkus2024}. For MAPPER, these studies justify treating
schema design as a methodological component of the workflow rather than as a
minor preprocessing detail.

\subsection{Gap addressed}
\label{subsec:rw_gap}

Across accessibility measurement, network-based methods, and open geospatial
workflow design, the recurring limitation is not the absence of usable metrics
but the absence of reproducible infrastructure that jointly provides
\begin{enumerate}
  \item multimodal travel-time graphs,
  \item harmonized opportunity inventories,
  \item continuous network-based accessibility fields, and
  \item reusable, analysis-ready outputs for downstream spatial modeling.
\end{enumerate}
Existing studies and data products satisfy subsets of these requirements, but
rarely all at once in a country-scale, open-data pipeline oriented toward
urban and regional systems analysis \citep{marwal2022, mazzulla2024,
nelson2019, fink2024, pfertner2023}. The documented algorithmic dependence of
accessibility results on tool internals \citep{higgins2022} adds a fifth face
to the same gap: reproducibility must extend through the routing and measure
computation itself, not only through the data assembly around it.

This gap is also substantively relevant for the Bavaria case study, where prior
planning work has called for stronger accessibility-based and timetable-based
assessment of regional service provision \citep{klein2022}.
MAPPER is positioned as a response to that
computational gap: not a digital twin, not a real-time system, and not a
single-purpose case study model, but a reproducible geocomputational workflow
for generating multimodal accessibility fields at national scale, on commodity hardware.

\section{Data}
\label{sec:data}

MAPPER uses open-data inputs for reproducible multimodal accessibility
analysis. At full-study scale, the workflow is demonstrated for Germany in
order to show country-scale computational feasibility. The resulting
accessibility dataset covers nine internal network and intermodal mode fields:
walk, bike, drive, transit, transit\_bus, transit\_rail, walk\_transit,
walk\_transit\_bus, and walk\_transit\_rail. All published outputs are sampled
to H3 resolution 10. This grid is fine enough to retain local spatial
gradients and act as a dense sampled representation of the underlying
continuous field, while remaining tractable for country-scale storage and
downstream joins. The Bavaria case study is then evaluated for four user-facing
modes: walk, bike, car, and public transport.

A representative Germany-wide production run makes the scale concrete. The
exported H3 table contains 39{,}992{,}878 cells and 972 accessibility columns,
corresponding to 108 base opportunity layers across nine modes, and occupies
about 66~GB in Parquet format. The exact column set depends on the configured
layer bundle, while the present paper focuses on the retained OSM- and
workplace-based layers and on the four modes used in the Bavaria application.

\subsection{Transport networks and temporal scope}
\label{subsec:transport_data}

The transport graphs are built from two open sources. Road and path networks
for walking, cycling, and car are derived from the Geofabrik Germany
OpenStreetMap (OSM) extract \citep{geofabrikGermanyExtract2025}. Public transport supply is
represented using the ``Germany Full'' GTFS feed published by \citet{brosiGermanyFullGtfs2026},
which is derived from the DELFI e.V. NeTEx dataset. The production run used the
feed version last updated on 2 May 2026 at 06:08:06 CEST, with approximately
1.6 million unique trips and 668{,}000 stops.

In the present study, public transport accessibility reflects scheduled service
for Monday departures within a morning analysis window from 07:00 to 09:00.
This window is used as a high-availability planning baseline: it captures
regular weekday service during a period when most routes are active, while
remaining simple enough for consistent national-scale comparison. The
GTFS-based layers should therefore be interpreted as timetable-based supply
under favorable scheduled conditions rather than as full-day or real-time
service quality.

\subsection{Opportunity layers and harmonization}
\label{subsec:opportunity_data}

Opportunity layers are derived primarily from OSM POIs and workplace proxies.
Raw OSM tags from namespaces such as \texttt{amenity}, \texttt{shop},
\texttt{tourism}, \texttt{leisure}, \texttt{healthcare}, \texttt{office}, and
\texttt{craft} are harmonized into a stable analytical schema of standardized
categories. The purpose of this schema is not to reproduce the full semantic
richness of OSM, but to create a transparent and reproducible opportunity
inventory that is stable enough for large-scale accessibility analysis.

Examples illustrate the logic. The standardized class \texttt{doctor} merges
\texttt{amenity=doctors}, \texttt{healthcare=doctor}, and
\texttt{office=physician}. \texttt{pharmacy} combines
\texttt{amenity=pharmacy}, \texttt{healthcare=pharmacy}, and
\texttt{shop=chemist}. \texttt{kindergarten} merges
\texttt{amenity=kindergarten}, \texttt{amenity=childcare}, and
\texttt{amenity=preschool}. Retail and leisure classes are treated similarly,
for example by aggregating multiple shop tags into standardized categories such
as \texttt{supermarket}, \texttt{convenience\_store}, or
\texttt{department\_store}. Some tags are excluded from the final opportunity
inventory when they are either too noisy for the present use case or better
treated as transport infrastructure rather than destinations; examples include
bench, parking, bus-station, and railway-station tags. The appendix reports the
schema structure and representative mappings
(Appendix~B).

This harmonization strategy is informed by prior work showing both the utility
of OSM-derived POIs for urban and travel-demand analysis and the need for
explicit aggregation and plausibility checks when building analytical taxonomies
\citep{klinkhardt2021, yang2022, yu2022, notelaers2024}. Because OSM's tagging
space is flexible and heterogeneous and lacks a single standardized ontology,
large-scale analysis commonly relies on explicit taxonomy construction or
ontology-backed semantic alignment to obtain stable, reusable categories
\citep{shbita2024, dsouza2021}. The YAML schema used here should be understood
in that sense: as a transparent, versioned taxonomy layer for reproducible
opportunity harmonization rather than as a claim that a universal POI standard
already exists inside OSM itself.

\subsection{Workplace proxies and census population}
\label{subsec:workplace_data}

Workplace opportunity layers are estimated from OSM-derived building and land-use
information using published proxy concepts that approximate employment
capacity from mapped usage classes, floor area, and activity type
\citep{klinkhardt2021, malkus2024, notelaers2024}. In this paper, these layers
are treated as accessibility-relevant proxy opportunities rather than as
official employment counts. Their main role is to provide a harmonized
large-area representation of workplace-related destinations within the same
open-data pipeline as the other opportunity layers.

Germany's 2022 Census is used for population-weighted aggregation on the H3
grid and in the Bavaria case study \citep{destatisZensusGridPopulation2022}. In this
study, census population is used primarily to weight resident
exposure and municipal fulfillment rather than to infer additional opportunity
layers. A brief plausibility check for the workplace proxies against official
Bavarian employment totals is reported in \autoref{subsec:wp_validation} and
Appendix~C.

Two caveats matter for interpretation. First, the workplace layers inherit the
limitations of open-data proxies and should therefore be interpreted as
approximate opportunity indicators. Second, the OSM schema is an analytical
harmonization layer rather than a claim that OSM provides a complete or
normatively correct classification of services. These uncertainties do not
invalidate the accessibility workflow, but they do narrow the kinds of
substantive claims that can be made from individual opportunity layers.

\section{Methods}
\label{sec:methods}

\subsection{Pipeline overview}
\label{subsec:methods_overview}

\begin{figure}[!htpb]
    \centering
    \includegraphics[width=0.85\textwidth]{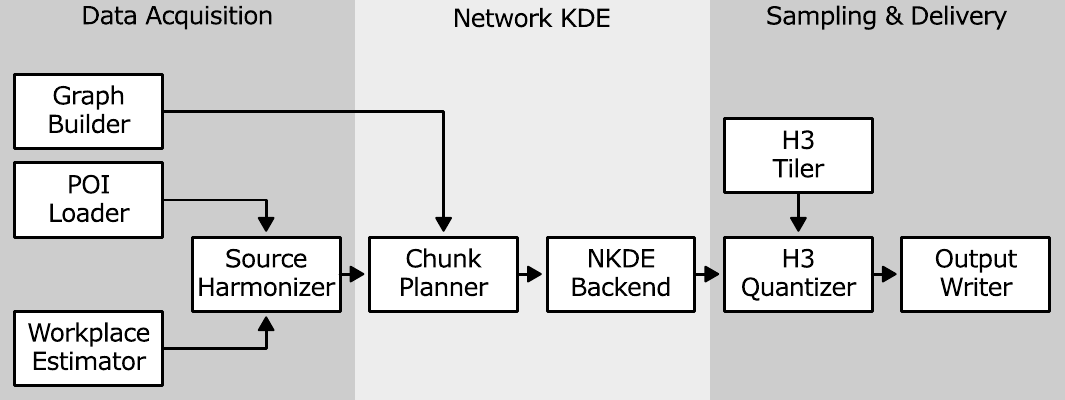}
    \caption{MAPPER pipeline architecture. Open geospatial inputs are converted
    into mode-specific travel-time graphs and harmonized opportunity layers,
    followed by NKDE-based accessibility inference and H3-aligned output
    generation.}
    \label{fig:mapper_pipeline_overview}
\end{figure}

As summarized in \autoref{fig:mapper_pipeline_overview}, the workflow consists
of three stages. First, open transport and opportunity data are ingested and
harmonized into mode-specific graphs and analysis-ready source layers. Second,
continuous accessibility fields are computed on the transport networks using
travel time as impedance. Third, the node-level results are sampled to an H3
grid and exported in a format that supports downstream geospatial analysis.

The production implementation is designed for national-scale runs rather than
interactive routing queries, and its computational architecture is a deliberate
part of the contribution. Existing accessibility tooling is typically consumed
as an interpreter-level package or a containerized service; this is convenient
and well suited to city-scale studies, but it layers orchestration and
per-query overhead on top of the routing core --- overhead that multiplies at
the scale targeted here, where the workload is the product of nine modes, more
than one hundred opportunity layers, and a national network. MAPPER instead
keeps the entire hot path in native code: shortest-path evaluation uses
customizable contraction hierarchies (RoutingKit), and the NKDE accumulation
and grid-sampling stages are implemented in C++ with OpenMP parallelism and
explicit memory management, so throughput scales with available cores from
workstation-class machines to many-core hosts. A DAG-style orchestration layer
schedules graph construction, source preparation, accessibility inference, and
sampling by data dependency, which keeps the critical path short and
systematically eliminates redundant computation --- shared travel-time
structures are built once per mode and reused across all opportunity layers
rather than recomputed per layer. These implementation choices affect runtime
and scalability rather than the conceptual definition of the accessibility
measure, but they are what make a national multimodal field stack practically
computable on commodity hardware.
Appendix~A summarizes the main implementation and
configuration parameters used in the production workflow and the Bavaria
application.

\subsection{Graph construction and source harmonization}
\label{subsec:source_harmonization}

MAPPER builds directed, time-weighted graphs for walking, cycling, car,
and scheduled public transport. For walk, bike, and car, edge travel times are
derived from network geometry and mode-specific speed assumptions. The
production configuration uses 5~km/h for walking. Cycling uses
highway-type-dependent speeds around a 19.5~km/h default, while car routing uses
OSM \texttt{maxspeed} values where available and road-class defaults otherwise;
car travel times are therefore free-flow estimates, and congestion is not
modeled. For public transport, GTFS schedules are converted into a
timetable-based graph and linked to the walking network at stop locations to
represent access, egress, and transfer movements within the selected analysis
window. Public transport routing uses earliest feasible departures in the
filtered Monday morning timetable; transfers between non-coincident stops are
represented by walking links. Reported public transport travel times start at
the departure of the first boarded service: initial waiting time before that
departure is not included, while waiting implied by timetable transfers within
the journey is. This convention is configurable in the workflow; under the
default used here it contributes to the deliberately supply-oriented,
optimistic character of the public transport estimates discussed in
Sections~\ref{sec:validation} and~\ref{sec:discussion}.

Opportunity layers are then mapped to a shared reference structure before
mode-specific accessibility is computed. POIs are assigned to nearby admissible
network nodes, while workplace proxies and other weighted layers are converted
into node-level source weights. This keeps the opportunity definition fixed
across modes, so modal differences arise from the travel-time structure of the
graphs rather than from inconsistent source placement.

\subsection{Accessibility model}
\label{subsec:nkde}

\begin{figure}[!htpb]
    \centering
    \includegraphics[width=\textwidth]{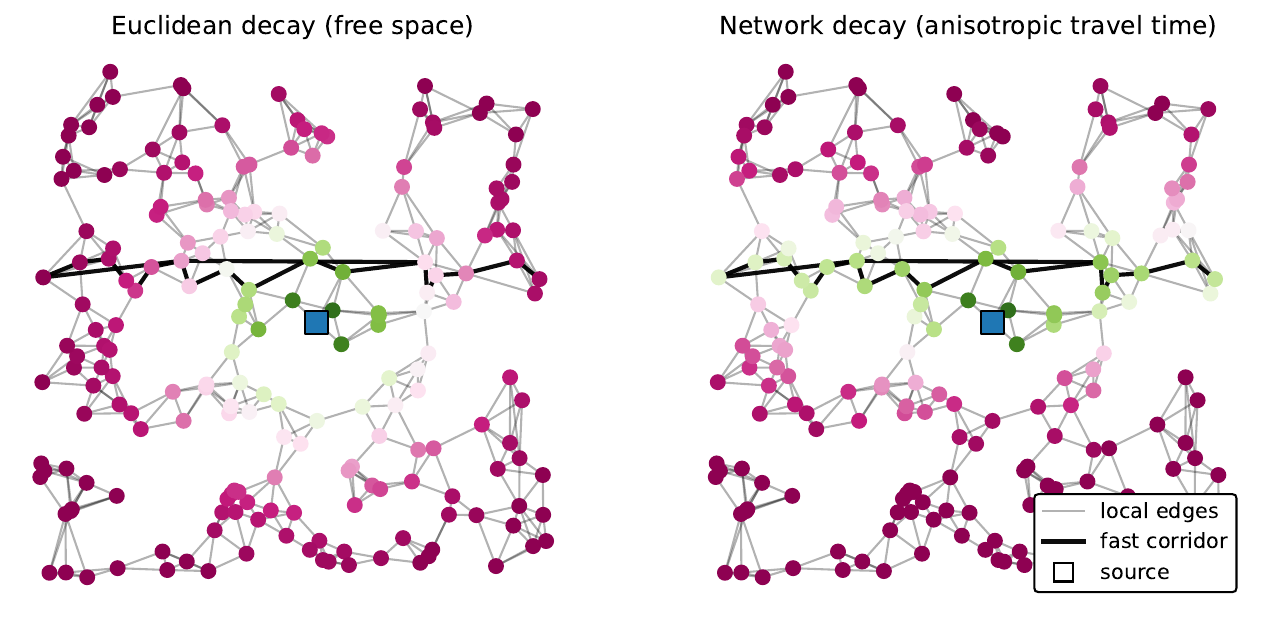}
    \caption{Euclidean versus network-constrained kernel smoothing. Euclidean smoothing distributes influence isotropically based on planar proximity, including into areas that are nearby in space but weakly connected in network travel time. The network-constrained formulation preserves corridor effects, barriers, and anisotropic reachability.}
    \label{fig:kde_comparison}
\end{figure}

For each mode $m$ and opportunity layer $\ell$, MAPPER computes a continuous
accessibility potential on the transport network using travel time as
impedance. \autoref{fig:kde_comparison} illustrates why a network-constrained
formulation is used instead of Euclidean smoothing: barriers and network
topology shape the resulting field directly rather than being blurred away by
planar distance.

Let $d_m(u,s)$ denote the shortest-path travel-time distance between query node
$u$ and source node $s$. Accessibility at node $u$ is computed as
\begin{equation}
\label{eq:nkde}
A_{m,\ell}(u) \;=\; \sum_{s\in S_{\ell,m}} w_{\ell}(s)\,K_h\!\bigl(d_m(u,s)\bigr),
\end{equation}
where $w_{\ell}(s)$ is the source weight and $K_h(\cdot)$ is a Gaussian kernel
over travel time:
\begin{equation}
\label{eq:gaussian_kernel}
K_h(t) = \exp\!\left(-\tfrac{1}{2}(t/h)^2\right).
\end{equation}
In this study, the kernel bandwidth is fixed at $h=850$~s ($\approx 14.2$~min).
This parameter governs the rate at which opportunity influence decays with
travel time and therefore the smoothness and effective reach of the resulting
accessibility fields. The Gaussian form was chosen because it yields smoothly
varying fields without hard discontinuities and is convenient for cross-mode
comparison, aggregation, and downstream modeling. Gaussian travel-time decay
forms have also been shown to align well with observed travel-time behavior in
gravity-style accessibility modeling, and Gaussian decay is widely used as a
continuous distance-decay representation in accessibility methods
\citep{santanapalacios2022, subal2021}.
The bandwidth value was chosen heuristically as a compromise between two
undesirable extremes: smaller values produce fragmented, highly local fields,
while larger values oversmooth the surface and allow dense urban clusters to
dominate too strongly over wider areas. A value of $850$~s preserves local
contrast and anisotropic network spread without excessive metropolitan
spillover. The cutoff of 7200~s is primarily a computational horizon and is set
sufficiently far from the selected bandwidth that the omitted Gaussian tail is
negligible for the reported analyses.

The resulting field values have a direct operational reading: a single
opportunity located exactly $\tau$ seconds away contributes $K_h(\tau)$ to the
field, so a value of $0.571$ is equivalent to one opportunity at 15 minutes
under the chosen bandwidth, and larger values indicate more and/or closer
opportunities. This is also precisely how the policy thresholds of the Bavaria
application are constructed
(Section~\ref{subsec:case_study}), which makes threshold choices on the
continuous field interpretable in travel-time terms rather than arbitrary.

\subsection{H3 sampling and output structure}
\label{subsec:h3_sampling}

\begin{figure}[!htpb]
    \centering
    \includegraphics[width=0.45\textwidth]{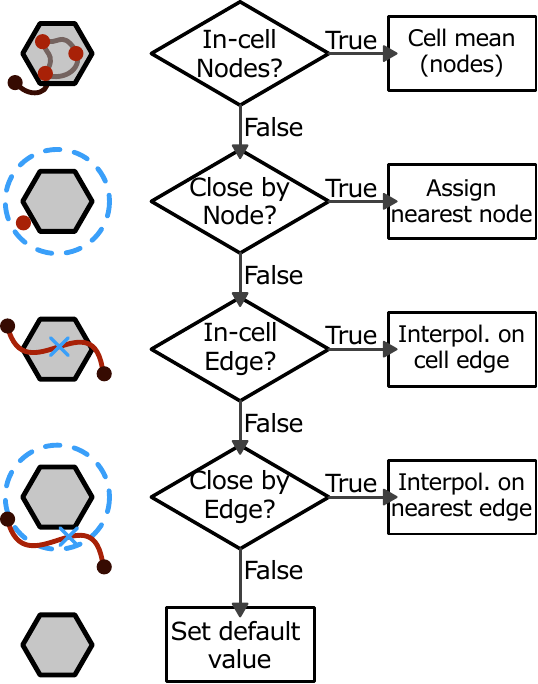}
    \caption{Hierarchical assignment of node-level accessibility values to H3
    cells.}
    \label{fig:sampling}
\end{figure}

Accessibility is computed on transport-network nodes, but the published outputs
are stored on a common H3 grid at resolution 10. As visible in
\autoref{fig:sampling}, node-level accessibility values are therefore assigned
to H3 cells using a hierarchical sampling policy so that the final layers can
be merged consistently across modes, themes, and external geospatial
covariates. The production sampler first uses graph nodes falling inside a
cell, then falls back to the nearest graph node within 100~m and finally to the
nearest eligible edge within 150~m; cells without a resolvable graph sample are
assigned the configured default value. These thresholds are reported with the
other run parameters in Appendix~A.

For each H3 cell $H_i$, MAPPER exports a feature vector
\[
\mathbf{z}_i = \bigl(z_{i,m,\ell}\bigr)_{m\in\mathcal{M},\,\ell\in\mathcal{L}},
\]
where $\mathcal{M}$ denotes the exported mode set and $\mathcal{L}$ the layer
set. This yields a country-scale matrix of multimodal accessibility features on
a uniform grid. The exported tables are keyed by H3 index and accompanied by
metadata on graph construction, opportunity schema version, analysis window,
and kernel settings.

\begin{figure}[!htpb]
    \centering
    \includegraphics[width=0.55\textwidth]{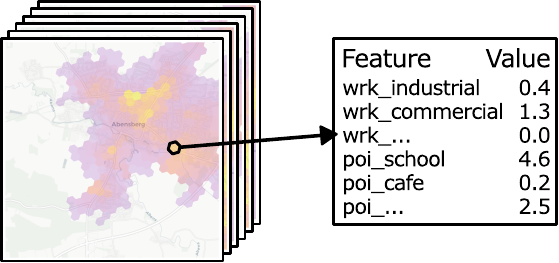}
    \caption{Structure of the exported MAPPER output bundle. Accessibility
    layers are written as H3-indexed columnar tables with accompanying metadata
    for reproducible downstream analysis.}
    \label{fig:output_bundle}
\end{figure}

This output design is intended to support reuse rather than a single reporting
workflow. The same accessibility stack can be joined to demographic layers,
used in map-based diagnostics, or passed to downstream statistical learning,
scenario, and optimization models without rebuilding the source data
foundation.

\section{Validation}
\label{sec:validation}

The routing diagnostics are interpreted as validation of the travel-time
impedance scale rather than full validation of accessibility inference. The
main purpose is to assess whether mode-specific travel times are approximately
unbiased in aggregate, because these values enter the NKDE directly. This
focus also targets exactly the layer where accessibility results have been
shown to diverge across tools --- the travel-time computation itself
\citep{higgins2022} --- which is why the validation addresses impedance
realism rather than downstream field values.

We compare predicted travel times against an empirical mobility panel containing
observed trip durations and mode information from smartphone-based travel-diary
studies in Germany \citep{loder2024, nachtigall2025}.
Origins and destinations are
snapped to the corresponding mode graphs using the same rules as in the
production workflow, and predicted travel time $t_{\text{pred}}$ is compared to
observed duration $t_{\text{obs}}$ using the signed error
\[
\Delta t = t_{\text{pred}} - t_{\text{obs}} .
\]
Signed error is the relevant quantity here because systematic offsets in travel
time translate directly into systematic shifts of the accessibility field.

We use two diagnostics. First, Bland--Altman plots summarize agreement and
mode-specific bias across the observed trip-duration range. Second, we examine
how the mean error behaves under aggregation, because MAPPER is used to produce
accessibility surfaces that pool information across many opportunities rather
than to predict individual route choice.

\section{Results}
\label{sec:results}

\subsection{Travel-time evaluation}
\label{subsec:traveltime_eval}

\begin{figure}[!htpb]
    \centering
    \includegraphics[width=\textwidth]{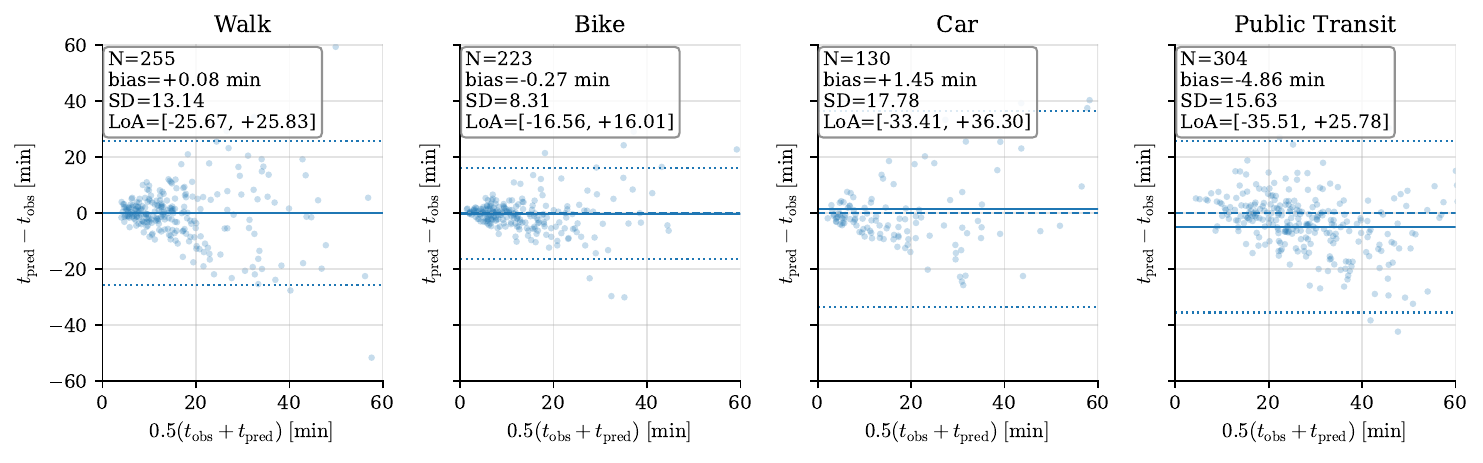}
    \caption{Bland--Altman plots for predicted versus observed travel times by
    mode. The y-axis shows $\Delta t=t_{\text{pred}}-t_{\text{obs}}$ and the
    x-axis shows the mean travel time $(t_{\text{pred}}+t_{\text{obs}})/2$.
    Solid line: mean bias. Dotted lines: limits of agreement (bias $\pm 1.96$
    SD).}
    \label{fig:ba_validation}
\end{figure}

\autoref{fig:ba_validation} shows small mean bias for walking, cycling, and
car, while public transport exhibits a negative bias. The sign of this bias is
expected under the routing convention described in
Section~\ref{subsec:source_harmonization}: reported journeys start at the
first boarded departure, so predicted times exclude the initial waiting that
observed door-to-door durations include. We interpret the public
transport pattern as an optimistic upper-bound effect rather than as evidence of
high timetable-routing fidelity: the graph represents scheduled service in a
high-availability morning window and does not capture missed connections,
reliability effects, or additional waiting outside the modeled timetable
conditions. This is consistent with \citet{wessel2019}, who find that
schedule-based GTFS accessibility systematically exceeds realized accessibility
by roughly 5--15\% in well-served areas and is more variable where service is
sparse. Public-transport-based accessibility is therefore better understood
as schedule-based supply under favorable assumptions than as realized
door-to-door travel time.

This matters for the case study because municipalities near the policy-derived
thresholds could switch pass/fail status under less favorable waiting or
transfer assumptions. The routing validation is therefore informative about the
plausibility of the impedance scale, but it does not demonstrate that threshold
classifications are insensitive to public transport schedule uncertainty. A
similar caveat applies to car, where uncongested travel times likely make car-based
accessibility look stronger than it would under peak congestion.

\begin{figure}[!htpb]
    \centering
    \includegraphics[width=\textwidth]{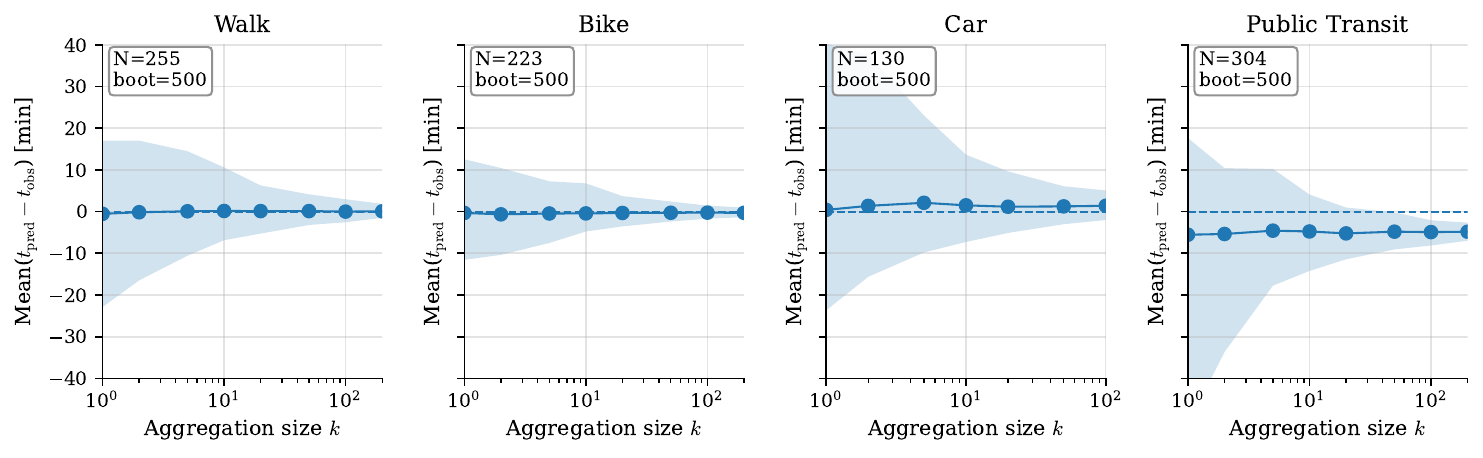}
    \caption{Aggregation stability of mean travel-time error. For each mode and
    aggregation size $k$, we sample $k$ trips with replacement and compute the
    mean error $\overline{\Delta t}_k$. Solid line: bootstrap mean. Shaded
    band: 95\% bootstrap interval.}
    \label{fig:mean_k_validation}
\end{figure}

As shown in \autoref{fig:mean_k_validation}, the uncertainty band around the
mean error narrows as aggregation size increases. This supports the use of the
graphs as a comparative impedance baseline for NKDE-based accessibility fields,
because those fields aggregate over many opportunity contributions. At the same
time, aggregation should not be read as fully resolving public-transport-specific
connection errors or other mode-specific biases. The safer interpretation is
that the routed graphs are adequate for comparative large-area accessibility
analysis, while fine-grained pass/fail statements near policy thresholds remain
sensitive to unmodeled timetable and congestion effects.

\subsection{Workplace proxy plausibility}
\label{subsec:wp_validation}

Because workplace-related opportunity layers are estimated from open geospatial
proxies rather than observed employment registers, we also perform a simple
scale check against official Bavarian sector totals from the Bavarian
statistics office's GENESIS-Online system \citep{bavariaGenesisEmploymentTable13312001r2026}.
Aggregated to three broad WZ2008 groups, the OSM-derived model recovers 91\% of
total employment, with a slight overestimate for industry and manufacturing
(ratio 1.08), an underestimate for services (0.85), and a larger underestimate
for agriculture (0.68). We interpret this as a
plausibility check rather than a full validation of workplace estimation: it
indicates that the proxy layers are of a reasonable order of magnitude for
accessibility analysis, while still inheriting expected OSM coverage biases and
classification gaps. The comparison details are reported in
Appendix~C.

\subsection{Bavaria case study operationalization}
\label{subsec:case_study}

The case study evaluates Bavaria's Central Places system using multimodal
accessibility rather than municipality-internal inventories alone. The
central-place system is a land-use planning instrument: designations assign
municipalities a role in the spatial distribution of services and thereby
presume a level of access that the transport system must deliver. The case
study asks exactly this transport--land-use question --- whether the access
presumed by the land-use plan is available within the plan's own travel-time
budgets, evaluated consistently across modes.
\citet{klein2022} argues that central-place designations in Bavaria require a
more explicit accessibility-based assessment, especially for public transport.
In the Bavarian \emph{Landesentwicklungsprogramm} (LEP), each municipality is
assigned a central-place level that corresponds to a different expected level
of service provision. In simplified terms, municipalities are designated as
basic, medium, or upper centers, and these designations imply different sets
of functions that should be accessible to residents within mode-specific
travel-time ranges. The planning question is therefore not only whether a given
service exists somewhere in the municipality, but whether the service bundle
associated with the municipality's assigned level is accessible within the
expected time budget. MAPPER makes it possible to evaluate this more directly
on a common spatial support across modes, whereas prior assessments often
relied more strongly on municipality-internal counts or other simplified
representations of provision. We therefore operationalize LEP service bundles
as thresholds on the continuous NKDE fields rather than as simple
within-boundary counts.

The required service indicators are grouped into subcategories for the three
central-place levels. For each inhabited H3 cell, subcategory fulfillment is
evaluated using the policy-derived thresholds from \autoref{eq:policy_activation}.
Need-level scores are then aggregated to
municipalities as population-weighted means. This procedure preserves
cross-border opportunity influence and supports like-for-like comparison across
walking, cycling, car, and schedule-based public transport.

The thresholding step is policy-derived and intentionally separate from the
kernel bandwidth itself. The bandwidth $h=850$~s defines how the continuous
accessibility field is constructed; the LEP-derived $\theta_{c,m}$ values are a
downstream classification device used only for the Bavaria application. This
distinction is important because the case study does not replace the NKDE with
hard travel-time cutoffs; it interprets the continuous field through
policy-relevant benchmarks. Recomputing the LEP thresholds with the same kernel
and bandwidth keeps the case study classification internally consistent with
the underlying NKDE field, but it does not remove bandwidth dependence because
the field still reflects cumulative contributions from multiple nearby
opportunities under the chosen decay scale. We verify in
Appendix~F that the case study's qualitative findings ---
and in particular the near-universal upper-level fulfillment for the motorised
modes --- are robust to this decay-scale dependence by re-deriving them with a
heavy-tailed exponential kernel whose long-horizon activation thresholds are
numerically non-degenerate.

More specifically, the LEP travel-time targets are translated back into NKDE
activation thresholds using the same Gaussian kernel rather than applied as
hard routing cutoffs. For a mode-specific policy target travel time
$\tau_{c,m}$, the corresponding classification threshold is
\begin{equation}
\label{eq:policy_activation}
\theta_{c,m} = K_h(\tau_{c,m}) =
\exp\!\left(-\tfrac{1}{2}(\tau_{c,m}/850)^2\right),
\end{equation}
with $\tau_{c,m}$ expressed in seconds. This is a policy-oriented
interpretation of the continuous accessibility field, not a one-to-one
nearest-destination travel-time test. The full set of LEP travel times and
derived activation thresholds used in the case study is reported in
Appendix~D.

\begin{figure}[!htpb]
    \centering
    \includegraphics[width=\textwidth]{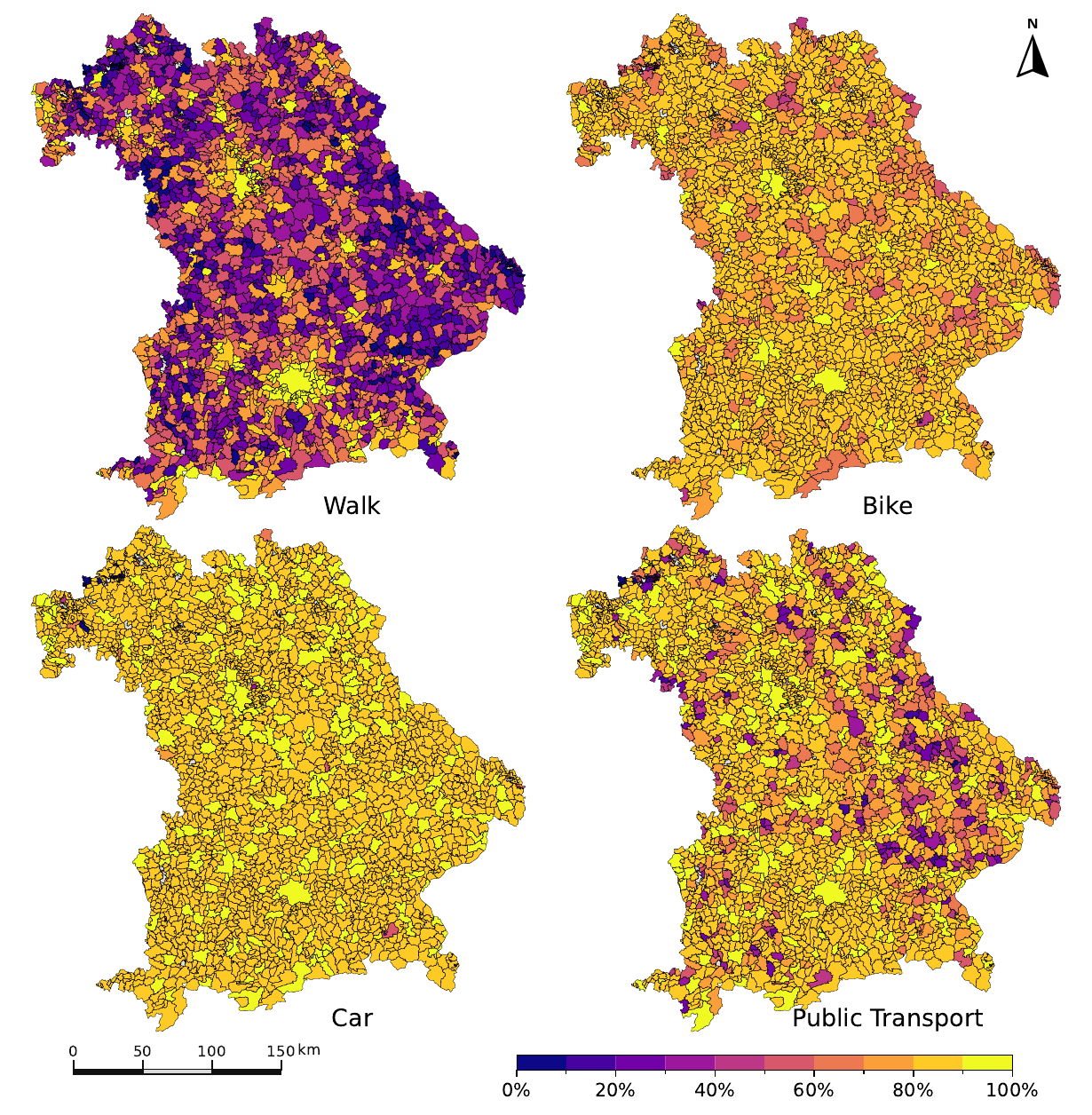}
    \caption{Population-weighted municipal fulfillment of assigned central-place
    functions in Bavaria by mode. For each municipality and mode, the mapped score
    is the population-weighted share of required service subcategories that meet
    the policy-derived NKDE thresholds, displayed as a percentage. Higher values
    indicate that a larger share of the assigned service requirements is accessible
    for the average resident.}

    \label{fig:bavaria_case_study}
\end{figure}

\autoref{fig:bavaria_case_study} provides the spatial context for the threshold-based summaries in \autoref{tab:muni_coverage_mean} and
\autoref{tab:pop_undersupply}. We evaluate three scenario thresholds on the
municipal fulfillment scores, with $T1$ as a strict cut-off ($\tau=0.90$),
$T2$ as a moderate cut-off ($\tau=0.60$), and $T3$ as a lenient cut-off
($\tau=0.30$). These thresholds are used as sensitivity scenarios rather than
as objective categories of adequacy. Appendix~E
shows the corresponding continuous threshold-response curves over
$\tau \in [0.10,0.90]$ and indicates that $T1$--$T3$ sample a broader
sensitivity continuum rather than representing isolated or arbitrary cut-offs.

\begin{table}[!htpb]
\centering
\small
\setlength{\tabcolsep}{5pt}
\renewcommand{\arraystretch}{1.15}
\caption{Shares by mode and target level. (M: municipalities meeting
requirement; P: population in municipalities meeting requirement.)}
\label{tab:muni_coverage_mean}
\begin{tabular}{llcccccc}
\toprule
\textbf{Mode} & \textbf{Level} & \multicolumn{2}{c}{\textbf{T1}} & \multicolumn{2}{c}{\textbf{T2}} & \multicolumn{2}{c}{\textbf{T3}} \\
\cmidrule(lr){3-4}\cmidrule(lr){5-6}\cmidrule(lr){7-8}
 &  & \textbf{M (\%)} & \textbf{P (\%)} & \textbf{M (\%)} & \textbf{P (\%)} & \textbf{M (\%)} & \textbf{P (\%)} \\
\midrule
\multirow{3}{*}{Walk} & basic & 4.4 & 10.2 & 20.1 & 34.6 & 61.7 & 75.8 \\
 & medium & 0.0 & 0.0 & 28.4 & 36.4 & 87.1 & 92.1 \\
 & upper & 0.0 & 0.0 & 73.8 & 93.5 & 98.4 & 100.0 \\
\midrule
\multirow{3}{*}{Bike} & basic & 92.3 & 96.1 & 99.6 & 100.0 & 99.9 & 100.0 \\
 & medium & 51.7 & 63.5 & 100.0 & 100.0 & 100.0 & 100.0 \\
 & upper & 82.0 & 96.6 & 100.0 & 100.0 & 100.0 & 100.0 \\
\midrule
\multirow{3}{*}{Car} & basic & 97.8 & 99.5 & 99.4 & 100.0 & 99.7 & 100.0 \\
 & medium & 98.5 & 99.0 & 100.0 & 100.0 & 100.0 & 100.0 \\
 & upper & 98.4 & 99.6 & 100.0 & 100.0 & 100.0 & 100.0 \\
\midrule
\multirow{3}{*}{\shortstack{Public\\transport}} & basic & 68.3 & 74.5 & 89.7 & 94.5 & 98.2 & 99.4 \\
 & medium & 70.1 & 75.9 & 99.0 & 99.4 & 100.0 & 100.0 \\
 & upper & 96.7 & 99.6 & 100.0 & 100.0 & 100.0 & 100.0 \\
\bottomrule
\end{tabular}
\end{table}

\autoref{tab:muni_coverage_mean} shows a stable modal ordering across the three
thresholds. Walking remains the limiting mode, especially under the strict and
moderate scenarios: under $T1$, only 4.4\% of municipalities and 10.2\% of the
population meet the basic LEP provisioning bundle on foot alone. By contrast,
car-based fulfillment is already near-universal under the strictest scenario,
with cycling and public transport in between. The ranking remains similar at
$T2$ and $T3$, even though overall coverage rises for all modes
(\autoref{tab:muni_coverage_mean}).

Two further patterns are visible. First, population coverage exceeds municipal
coverage for almost all combinations of mode, level, and threshold. Shortfalls
therefore concentrate disproportionately in smaller or less populous
municipalities. Second, assigned upper-level municipalities tend to perform
better than lower-level municipalities under the same modal scenario, which is
consistent with the concentration of higher-order centers in denser and better
connected parts of the state. This should not be read as proof that the target
system is normatively well calibrated; it indicates that the current
designation pattern is associated with stronger accessibility outcomes in the
more connected urban parts of Bavaria.

\begin{table}[!htpb]
\centering
\small
\setlength\tabcolsep{3pt}
\renewcommand{\arraystretch}{1.15}
\sisetup{detect-all=true}
\begin{threeparttable}
\caption{Population exposed to undersupply by threshold and transport mode
(shares below $\tau$).}
\label{tab:pop_undersupply}
\begin{tabular}{l S[table-format=1.2,table-number-alignment=center] S[table-format=3.1,table-number-alignment=center] S[table-format=3.1,table-number-alignment=center] S[table-format=3.1,table-number-alignment=center] S[table-format=3.1,table-number-alignment=center] S[table-format=3.1,table-number-alignment=center] S[table-format=3.1,table-number-alignment=center] S[table-format=3.1,table-number-alignment=center] S[table-format=3.1,table-number-alignment=center]}
\toprule
\multicolumn{1}{c}{\textbf{Threshold}} & \multicolumn{1}{c}{\textbf{$\tau$}} & \multicolumn{2}{c}{\textbf{Walk}} & \multicolumn{2}{c}{\textbf{Bike}} & \multicolumn{2}{c}{\textbf{Car}} & \multicolumn{2}{c}{\textbf{\shortstack{Public\\transport}}} \\
\cmidrule(lr){3-4} \cmidrule(lr){5-6} \cmidrule(lr){7-8} \cmidrule(lr){9-10}
\multicolumn{1}{l}{} & \multicolumn{1}{c}{} & \multicolumn{1}{c}{BU} & \multicolumn{1}{c}{TU} & \multicolumn{1}{c}{BU} & \multicolumn{1}{c}{TU} & \multicolumn{1}{c}{BU} & \multicolumn{1}{c}{TU} & \multicolumn{1}{c}{BU} & \multicolumn{1}{c}{TU} \\
\midrule
T1 & 0.90 & 46.5 & 68.7 & 2.9 & 9.6 & 0.3 & 0.3 & 6.5 & 6.6 \\
T2 & 0.60 & 24.6 & 28.1 & 0.2 & 0.2 & 0.3 & 0.3 & 3.6 & 3.6 \\
T3 & 0.30 & 14.1 & 14.9 & 0.2 & 0.2 & 0.3 & 0.3 & 1.7 & 1.6 \\
\bottomrule
\end{tabular}
\begin{tablenotes}[flushleft]
\footnotesize
\item Notes: Values are population shares in \% below threshold $\tau$. BU =
Basic unmet (uses basic score everywhere). TU = Target unmet (uses the
municipality's assigned LEP target level score). Because TU may use a
different score than BU, TU can be lower or higher than BU.
\end{tablenotes}
\end{threeparttable}
\end{table}

\autoref{tab:pop_undersupply} complements the municipal perspective with a
resident-exposure perspective. The same broad mode ordering reappears, but the
contrast between territorial and population-based summaries becomes more
explicit. Car exposure to undersupply remains negligible across all three
thresholds, cycling exposure is low once the threshold is relaxed beyond $T1$,
and public transport exposure drops from 6.5--6.6\% at $T1$ to around 1.6--3.6
\% at $T2$--$T3$. Walking remains the outlier: even at the moderate threshold,
24.6\% of residents fall below the basic LEP provisioning bundle and 28.1\%
fall below the target-specific LEP requirement.

The difference between BU and TU is especially relevant for walking at $T1$.
Using the municipality's assigned target level raises the exposed population
from 46.5\% to 68.7\%, which indicates that stricter assigned requirements are
not evenly supported by walking accessibility. For bike, car, and public
transport, the BU--TU difference is much smaller. This does not imply that the
LEP classifications are invalid or validated by the analysis; it shows how the
same MAPPER field can be used to compare alternative interpretations of service
adequacy under a consistent multimodal accessibility model.

\section{Discussion}
\label{sec:discussion}

\paragraph{What the case study shows}
The Bavaria application demonstrates that the same accessibility field can
support multiple policy-oriented diagnostics without reverting to
within-boundary inventories. Cross-border opportunity influence is preserved,
all modes are evaluated on the same spatial support, and municipal summaries
can be paired with resident-exposure summaries. Under this operationalization,
the modal ordering is robust across thresholds: walking remains the most
restrictive mode, car the least restrictive, and cycling and public transport
occupy intermediate positions. For the transport--land-use question the
central-place system embodies, this modal ordering is itself informative: the
service levels presumed by the land-use designations are consistently easier
to reach by car than by the sustainable modes the plan also aims to support,
and the size of that gap varies systematically across space. The analysis
therefore supports comparative interpretation, but not a categorical judgment
that the Bavarian central-place system is either ``working'' or ``failing'' in
any absolute sense.

\paragraph{What MAPPER enables}
The case study's contribution is methodological rather than normative: it
demonstrates a transferable way of operationalizing statutory planning targets
on continuous accessibility fields. MAPPER produces continuous, harmonized,
multimodal accessibility layers that can be translated into different
planning-oriented score systems while retaining a consistent computational
backbone. This makes
it possible to move between map-based diagnostics, municipal summaries,
population-weighted exposure measures, and alternative scenario thresholds
without rebuilding the data foundation each time. In that sense, the Bavaria
example is a showcase of reuse: the paper's contribution is the geospatial
analytics framework and aligned output structure, while the LEP-oriented
application illustrates one class of downstream analysis that the framework can
support. The same output structure is also suitable for downstream statistical
learning and scenario-oriented geospatial modeling, because it preserves dense,
mode-aligned accessibility features on a common spatial support.

\paragraph{Uncertainty and limits of interpretation}
Several limitations remain important for interpretation. First, the case study
translates LEP time budgets into NKDE activation thresholds rather than testing
direct nearest-destination compliance, so the results should be read as
policy-oriented classifications on a continuous accessibility field. Second,
the resulting scores depend on kernel bandwidth, threshold choice, and the
final opportunity schema; they are therefore scenario-dependent rather than
parameter-free facts. Third, public transport results remain optimistic
schedule-based estimates and car results likely overstate practical
accessibility because congestion is not modeled. Finally, the thematic schema
itself is a harmonized analytical construct derived from open geospatial data,
not an error-free census of service provision. These caveats narrow the policy
claims that can be made, but they do not weaken the central computational
point: reproducible multimodal accessibility fields make such assumptions
visible and comparable instead of burying them inside opaque inventories or
ad hoc local workflows.

\section{Conclusion}
\label{sec:conclusion}

This paper presented MAPPER as a reproducible open-data workflow for
large-area multimodal accessibility analysis. By combining OSM- and
GTFS-derived travel-time graphs, harmonized opportunity layers, NKDE-based
accessibility inference, and H3-aligned outputs, the workflow produces a
consistent accessibility feature stack that can be reused across urban and
regional analysis tasks rather than rebuilt for each case study.

The Germany-wide run demonstrates country-scale computational feasibility,
while the Bavaria application shows how the resulting fields can support
comparative planning diagnostics across modes, thresholds, and summary
representations. The validation results support the use of the routed graphs as
an impedance baseline for aggregated accessibility analysis, and the workplace
plausibility check provides an additional scale check for one of the main proxy
input families. At the same time, the case study interpretation remains
conditional on timetable assumptions, threshold choices, schema decisions, and
the absence of congestion.

The main contribution is therefore not a single policy verdict for Bavaria, but
a reproducible framework that makes multimodal accessibility analysis more
transparent, comparable, and reusable at national scale. For transport and
land-use research, consistent national accessibility fields enable a class of
analyses that previously required per-study data assembly: plan-consistency
checks of the kind demonstrated here for the central-place system,
cross-regional comparisons of modal access gaps, and scenario analysis of
service or land-use changes on a fixed computational backbone. Future work
should focus on deeper robustness diagnostics, richer temporal dynamics,
coupling the accessibility fields to land-use change and service-planning
scenarios, and stronger integration with downstream scenario and optimization
workflows.

\section*{Code and data availability}
The MAPPER source code is openly available under the Apache License 2.0 at
\url{https://github.com/TUMFTM/MAPPER}; this paper uses release v2.0.0
(\url{https://doi.org/10.5281/zenodo.23557362}). Configuration examples, the
OSM opportunity schema, installation instructions, and reproduction materials are included with the
code. The data are described in Section~\ref{sec:data} and Appendix~A in sufficient detail to identify
and obtain the same inputs.

\section*{Acknowledgments}
The authors disclose that generative AI tools were used to prepare this paper.
Specifically, GPT-5 was used for editing language and improving readability. The
authors have reviewed this and take full responsibility for the content of this
paper.

This research was funded by the Federal Ministry of Research, Technology and Space (BMFTR) as part
of the M-Cube Cluster (Clusters4Future) within the project DatSim2.0 (grant no.\ 03ZU2105HA). The
funder had no role in study design; data collection, analysis, or interpretation; manuscript
writing; or the decision to publish.

\section*{Declaration of competing interest}
The authors declare that they have no competing interests.

\printbibliography

\clearpage
\appendix
\setcounter{section}{0}
\renewcommand{\thesection}{\Alph{section}}
\makeatletter
\@addtoreset{equation}{section}
\@addtoreset{table}{section}
\@addtoreset{figure}{section}
\makeatother
\renewcommand{\theequation}{\thesection.\arabic{equation}}
\renewcommand{\thetable}{\thesection.\arabic{table}}
\renewcommand{\thefigure}{\thesection.\arabic{figure}}

\section{Selected Workflow Parameters}
\label{app:workflow_parameters}

\autoref{tab:workflow_parameters} summarizes the main production parameters
that shape input selection, graph construction, public-transport routing, NKDE,
H3 sampling, and output generation in the current workflow. These values are
reported here to keep the main methods section readable while still documenting
the implementation settings that matter for reproducibility.

\begingroup
\footnotesize
\setlength{\LTpre}{0pt}
\setlength{\LTpost}{0pt}
\begin{longtable}{p{0.29\textwidth}p{0.23\textwidth}p{0.40\textwidth}}
\caption{Selected reproducibility parameters used in the production pipeline
and the Bavaria case study.}
\label{tab:workflow_parameters}\\
\toprule
\textbf{Parameter} & \textbf{Value} & \textbf{Role} \\
\midrule
\endfirsthead
\caption[]{Selected reproducibility parameters used in the production pipeline
and the Bavaria case study (continued).}\\
\toprule
\textbf{Parameter} & \textbf{Value} & \textbf{Role} \\
\midrule
\endhead
\bottomrule
\endfoot
\multicolumn{3}{l}{\textit{Input data versions}} \\
OSM source & Geofabrik Germany extract & Road, path, and OSM-derived opportunity source; the exact extract/access date is recorded in the data citation and run manifest \\
GTFS feed & Germany Full & National public-transport feed from the DELFI e.V. NeTEx dataset \citep{brosiGermanyFullGtfs2026} \\
GTFS feed version & Last update 2026-05-02 06:08:06 CEST & Timetable version used for the production run \\
GTFS feed size & 239 MB; 1.6M unique trips; 668k stops & Feed-scale metadata reported by the feed provider \\
GTFS license & Creative Commons 4.0 & License reported by the feed provider \\
\midrule
\multicolumn{3}{l}{\textit{Modes and graph construction}} \\
Germany-wide exported internal mode fields & walk, bike, drive, transit, transit\_bus, transit\_rail, walk\_transit, walk\_transit\_bus, walk\_transit\_rail & Full mode set used in the national output bundle \\
Bavaria case study modes & walk, bike, car, public transport & Subset used in the reported policy application \\
Walk speed & 5 km/h & Constant speed used for walking edges and pedestrian access/egress links \\
Bike speed assumptions & Highway-type defaults around 19.5 km/h & Cycling edge times derived from network geometry and highway-type speed defaults \\
Car speed assumptions & OSM \texttt{maxspeed}; road-class defaults where missing & Car edge times represent uncongested network travel time \\
\midrule
\multicolumn{3}{l}{\textit{Public transport timetable handling}} \\
Service day & Monday & GTFS feed filtered to Monday service before accessibility computation \\
GTFS analysis window & 07:00--09:00 & Departures within the morning window used as high-availability scheduled supply \\
Public transport routing objective & Earliest feasible departure & Timetable graph represents scheduled departures in the filtered window rather than full-day average service \\
Walk--public-transport connector distance & 300 m & Maximum distance for linking stops to the walking graph \\
Public transport transfer logic & Walking link for non-coincident transfer nodes & Transfers between nearby but non-identical stop nodes include walking time; coincident nodes transfer directly \\
\midrule
\multicolumn{3}{l}{\textit{Opportunity snapping and NKDE}} \\
Kernel type & Gaussian & Distance-decay function used in NKDE \\
Kernel bandwidth $h$ & 850 s & Governs smoothing and effective reach of accessibility fields \\
NKDE cutoff & 7200 s & Maximum travel-time horizon considered during accumulation \\
Source snap mode & node & Assigns opportunities to admissible graph nodes before NKDE \\
Maximum snap distance & 200 m & Discards sources that are too far from the reference graph \\
\midrule
\multicolumn{3}{l}{\textit{H3 sampling and export}} \\
H3 resolution & 10 & Common output grid for all exported fields \\
H3 node fallback threshold & 100 m & Maximum distance from cell centroid to nearest graph node when no node falls inside the cell \\
H3 edge fallback threshold & 150 m & Maximum distance to nearest eligible graph edge when node fallback is unavailable \\
Default unsampled-cell value & 0 & Value assigned when no node or edge sample can be resolved within the configured thresholds \\
Export key & H3 index & Primary key used for all published accessibility tables \\
Exported accessibility schema & \texttt{\{layer\}\_\{mode\}} columns & One numeric column per opportunity layer and mode, plus metadata describing modes, layers, kernel settings, graph construction, and schema version \\
Run metadata schema & schema version 1 & Versioned run manifest accompanying the exported H3 tables \\
Routing backend & C++ pipeline with RoutingKit CCH & National-scale shortest-path evaluation\\
\end{longtable}
\endgroup

\section{OSM Opportunity Schema}
\label{app:osm_schema}

This appendix summarizes the analytical OSM harmonization used in MAPPER. Raw
OSM tags are grouped into standardized classes before accessibility inference.
The mapping is intended to stabilize noisy source taxonomies for analysis, not
to reproduce the full OSM ontology. \autoref{tab:osm_schema_summary_a} and
\autoref{tab:osm_schema_summary_b} report the main schema families and
representative class mappings; the implementation
additionally applies priority rules and excludes selected classes such as bus stations,
railway stations, parking, and benches from the final opportunity inventory.

\begin{table}[!htpb]
\centering
\footnotesize
\caption{Summary of the OSM opportunity harmonization used in MAPPER
(families 1--5).}
\label{tab:osm_schema_summary_a}
\begin{tabularx}{\textwidth}{p{0.2\textwidth}XX}
\toprule
\textbf{Family} & \textbf{Representative standardized classes} & \textbf{Example merged source tags} \\
\midrule
Healthcare and emergency & hospital, doctor, clinic, dentist, pharmacy, veterinary, police, fire\_station, ambulance\_station & doctor $\leftarrow$ amenity=doctors, healthcare=doctor, office=physician; pharmacy $\leftarrow$ amenity=pharmacy, healthcare=pharmacy, shop=chemist \\
Food and drink & restaurant, fast\_food, cafe, bar, pub, ice\_cream, biergarten, food\_court, drinking\_water & cafe $\leftarrow$ amenity=cafe, shop=coffee; drinking\_water $\leftarrow$ amenity=drinking\_water, man\_made=water\_tap \\
Education & school, university, kindergarten, library, music\_school, driving\_school & kindergarten $\leftarrow$ amenity=kindergarten, amenity=childcare, amenity=preschool; university $\leftarrow$ amenity=university, amenity=college \\
Finance & bank, atm, bureau\_de\_change & atm $\leftarrow$ amenity=atm, amenity=payment\_terminal \\
Mobility services & fuel\_station, bicycle\_parking, ferry\_terminal, charging\_station, taxi\_stand, car\_rental, bicycle\_rental & fuel\_station $\leftarrow$ amenity=fuel; taxi\_stand $\leftarrow$ amenity=taxi \\
\bottomrule
\end{tabularx}
\end{table}

\begin{table}[!htpb]
\centering
\footnotesize
\caption{Summary of the OSM opportunity harmonization used in MAPPER
(families 6--10).}
\label{tab:osm_schema_summary_b}
\begin{tabularx}{\textwidth}{p{0.2\textwidth}XX}
\toprule
\textbf{Family} & \textbf{Representative standardized classes} & \textbf{Example merged source tags} \\
\midrule
Accommodation & hotel, motel, guest\_house, hostel, campsite & campsite $\leftarrow$ tourism=camp\_site, tourism=caravan\_site \\
Retail & supermarket, convenience\_store, bakery, clothing\_store, department\_store, electronics, hardware\_store, book\_store, sports\_shop, hairdresser, butcher, greengrocer & clothing\_store $\leftarrow$ shop=clothes, shop=fashion, shop=boutique; department\_store $\leftarrow$ shop=department\_store, shop=mall \\
Leisure and culture & park, playground, sports\_centre, swimming\_pool, stadium, pitch, museum, cinema, theatre, arts\_centre, nightclub, zoo, viewpoint, attraction, picnic\_site, dog\_park & sports\_centre $\leftarrow$ leisure=sports\_centre, leisure=fitness\_centre; park $\leftarrow$ leisure=park, landuse=village\_green \\
Public services & post\_office, town\_hall, community\_centre, place\_of\_worship, toilets, recycling, waste\_disposal, shelter, prison & town\_hall $\leftarrow$ amenity=townhall, amenity=courthouse; community\_centre $\leftarrow$ amenity=community\_centre, amenity=social\_facility \\
Craft and industry & brewery, winery, workshop & workshop $\leftarrow$ craft=carpenter, craft=shoemaker, craft=electrician, craft=plumber \\
\bottomrule
\end{tabularx}
\end{table}

\section{Workplace Proxy Plausibility Check}
\label{app:wp_validation}

This appendix reports the aggregate plausibility check for the workplace proxy
layers used in MAPPER. The benchmark values are official Bavarian employment
totals from the GENESIS-Online database of the Bavarian statistics office
\citep{bavariaGenesisEmploymentTable13312001r2026}. The OSM-derived estimates are
aggregated to broad WZ2008 sector groups in order to test whether the proxy
layers are strongly mis-scaled before they are used in accessibility analysis.

\begin{table}[!htpb]
\centering
\small
\caption{Comparison of official Bavarian employment totals and OSM-derived
workplace proxy estimates by broad sector group.}
\label{tab:wp_validation}
\begin{tabular}{lrrr}
\toprule
\textbf{Sector group} & \textbf{Official} & \textbf{Model} & \textbf{Ratio} \\
\midrule
A: Agriculture & 110{,}100 & 74{,}603 & 0.68 \\
B--F: Industry and manufacturing & 2{,}068{,}400 & 2{,}230{,}485 & 1.08 \\
G--T: Services & 5{,}691{,}500 & 4{,}844{,}206 & 0.85 \\
Total & 7{,}870{,}000 & 7{,}149{,}294 & 0.91 \\
\bottomrule
\end{tabular}
\end{table}

The largest deviation occurs for agriculture, where greenhouse footprints are
an incomplete proxy for actual employment. Industry and manufacturing are close
to parity after applying single-storey defaults to ground-floor building types,
while services remain somewhat underestimated because a substantial share of
service employment is embedded in mixed-use or generically tagged buildings.
For the purposes of this paper, these deviations support a cautious use of the
workplace layers as proxy opportunity fields rather than as calibrated
employment counts.

\section{LEP Threshold Translation for the Bavaria Case Study}
\label{app:policy_thresholds}

This appendix reports the travel-time targets taken from the Bavarian
\emph{Landesentwicklungsprogramm} (LEP)
\citep{bavariaLepReadingVersion2023,bavariaLepAnhangZentraleOrte2023} and the corresponding NKDE activation
thresholds derived with the policy-activation mapping defined in the main text. These values are used
only in the Bavaria case study and are not part of the core MAPPER field
generation workflow.

\begin{table}[!htpb]
\centering
\small
\caption{LEP travel-time targets and corresponding NKDE activation thresholds
used in the Bavaria case study.}
\label{tab:policy_thresholds}
\begin{tabular}{llcc}
\toprule
\textbf{Level} & \textbf{Mode} & \textbf{Travel time (min)} & \textbf{$\theta_{c,m}$} \\
\midrule
\multirow{4}{*}{Basic} & Walk & 15 & 0.571 \\
 & Bike & 20 & 0.369 \\
 & Car & 20 & 0.369 \\
 & Public transport & 30 & 0.106 \\
\midrule
\multirow{4}{*}{Medium} & Walk & 20 & 0.369 \\
 & Bike & 25 & 0.211 \\
 & Car & 20 & 0.369 \\
 & Public transport & 45 & $6.44\times10^{-3}$ \\
\midrule
\multirow{4}{*}{Upper} & Walk & 25 & 0.211 \\
 & Bike & 30 & 0.106 \\
 & Car & 60 & $1.27\times10^{-4}$ \\
 & Public transport & 90 & $1.72\times10^{-9}$ \\
\bottomrule
\end{tabular}
\end{table}

\section{Threshold Sensitivity of Population Exposure}
\label{app:threshold_sensitivity}

\begin{figure}[!htpb]
    \centering
    \includegraphics[width=\textwidth]{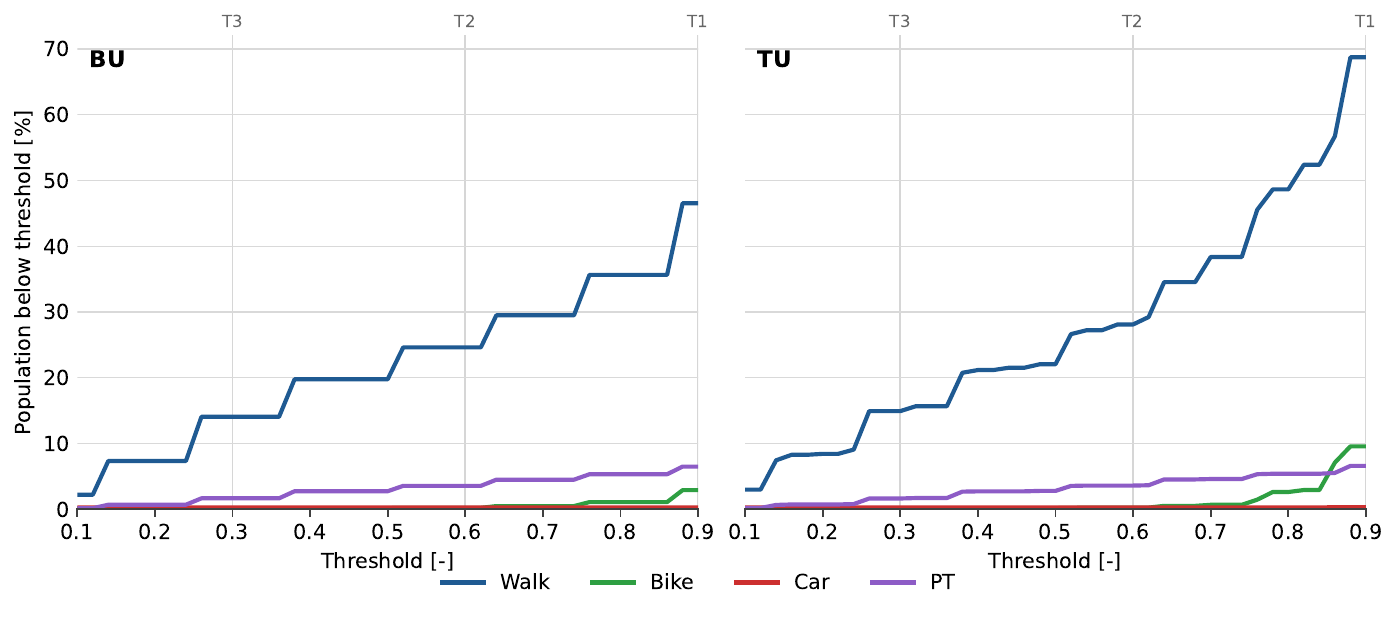}
    \caption{Continuous threshold sensitivity of population exposure to
    undersupply. Left: BU, where all cells are evaluated against the
    basic-level score. Right: TU, where cells are evaluated against the score
    implied by the municipality's assigned target level. The x-axis reports the
    classification threshold $[\,-\,]$, the y-axis reports the population
    below threshold [\%], and the vertical guide lines indicate the three
    scenario thresholds used in the main text ($T3=0.30$, $T2=0.60$,
    $T1=0.90$).}
    \label{fig:threshold_sensitivity_appendix}
\end{figure}

\autoref{fig:threshold_sensitivity_appendix} reports the continuous threshold
sensitivity of the population-exposure summary used in
the population-exposure table of the main text. The left panel shows the share of population
below threshold under the basic unmet interpretation (BU), where all cells are
evaluated against the basic-level score. The right panel shows the target unmet
interpretation (TU), where each cell is evaluated against the score implied by
its municipality's assigned central-place level. The vertical guide lines mark
the three scenario thresholds used in the main text ($T3=0.30$, $T2=0.60$,
$T1=0.90$).

The figure shows that the broad modal ordering is stable across the threshold
range: walking remains the most restrictive mode, car remains least
restrictive, and cycling and public transport occupy intermediate positions.
The TU curves are generally above the BU curves, especially for walking at
high thresholds, indicating that stricter assigned target levels increase the
share of population counted as undersupplied. This supports the interpretation
of $T1$--$T3$ as transparent sensitivity anchors on a continuous spectrum
rather than as uniquely privileged cut-offs.

\section{Kernel Sensitivity of the Central-Place Fulfillment}
\label{app:kernel_sensitivity}

The activation thresholds in \autoref{app:policy_thresholds} are derived from
the Gaussian kernel used to construct the accessibility fields
(the policy-activation mapping defined in the main text). Because the Gaussian decays rapidly, the
thresholds for the longest policy travel times are numerically very small: the
upper-level public-transport target ($90$~min) maps to
$\theta \approx 1.7\times10^{-9}$ and the upper-level car target ($60$~min) to
$\theta \approx 1.3\times10^{-4}$ (\autoref{tab:policy_thresholds}). A threshold
this close to zero effectively tests for the presence of \emph{any} reachable
opportunity rather than for reachability within the stated time budget, which
raises the question of whether the near-universal upper-level fulfillment
reported for the motorised modes is an artefact of the threshold rather than a
substantive result.

To test this, we recomputed the Bavarian accessibility fields with an
exponential distance-decay kernel, $K(t)=\exp(-\beta t)$ with
$\beta=10^{-3}\,\mathrm{s}^{-1}$ (a decay constant of $1000$~s, close to the
$\approx 850$~s scale of the Gaussian), holding all other settings fixed and
re-deriving the thresholds with the same kernel. The exponential tail keeps the
long-horizon thresholds numerically well resolved --- the $90$~min target now
maps to $\theta = 4.5\times10^{-3}$ and the $60$~min target to
$\theta = 2.7\times10^{-2}$, six and two orders of magnitude above their
Gaussian counterparts (\autoref{tab:kernel_thresholds}).

Under the exponential kernel the qualitative findings of the case study
persist: walking remains by far the most restrictive mode, motorised and cycling
access is broadly available, and coverage rises monotonically from the strict to
the lenient scenario. Crucially, the near-universal upper-level fulfillment for
car and public transport remains near-universal even against the numerically
non-degenerate exponential thresholds. We therefore conclude that this result
reflects genuinely broad motorised reachability of the upper-level service
bundle across Bavaria, not an artefact of the small Gaussian long-horizon
thresholds.

\begin{table}[!htpb]
\centering
\small
\caption{LEP activation thresholds under the Gaussian kernel used for the main
analysis and under the exponential kernel used for the robustness check.}
\label{tab:kernel_thresholds}
\begin{tabular}{llcc}
\toprule
\textbf{Level} & \textbf{Mode (travel time)} & \textbf{Gaussian $\theta$} & \textbf{Exponential $\theta$} \\
\midrule
Basic  & Walk ($15$~min)          & $0.571$              & $0.407$ \\
       & Public transport ($30$~min) & $0.106$           & $0.165$ \\
\midrule
Medium & Public transport ($45$~min) & $6.4\times10^{-3}$ & $0.067$ \\
\midrule
Upper  & Car ($60$~min)           & $1.3\times10^{-4}$   & $0.027$ \\
       & Public transport ($90$~min) & $1.7\times10^{-9}$ & $4.5\times10^{-3}$ \\
\bottomrule
\end{tabular}
\end{table}

The absolute fulfillment shares do shift between the two kernels, and the
exponential kernel is \emph{not} adopted for the main analysis, for a specific
reason. The NKDE field is a distance-decay-weighted \emph{sum} over all
reachable opportunities, whereas the activation threshold represents a single
opportunity at the target travel time. The heavy exponential tail therefore
lets high-catchment modes accumulate contributions from a large number of
distant opportunities, so the summed field exceeds the single-opportunity
threshold by one to three orders of magnitude. For a representative Bavarian
cell, the median non-zero supermarket field is $\approx 7.5$ by public transport
and $\approx 118$ by car, against thresholds of $0.17$ and $0.30$ respectively,
so the fulfillment test saturates and loses descriptive resolution. Walking,
whose field is dominated by a handful of nearby opportunities (median
$\approx 0.45$, against a threshold of $0.41$), stays discriminating under both
kernels. The faster-decaying Gaussian keeps the field local and thus preserves
more granularity in the fulfillment shares; it is retained for the main results,
with the exponential recomputation serving only as the robustness check reported
here.

\end{document}